\documentstyle[12pt,aaspp4,psfig]{article}

\def\ale{\mathrel{\hbox{\rlap{\hbox{\lower4pt\hbox{$\sim$}}}\hbox{$<$}}}}
\def\age{\mathrel{\hbox{\rlap{\hbox{\lower4pt\hbox{$\sim$}}}\hbox{$>$}}}}
\def\arcmin{\hbox{$^\prime$}}
\def\arcsec{\hbox{$^{\prime\prime}$}}

\def\gsim{\mathrel{\hbox{\rlap{\lower.55ex \hbox {$\sim$}}
                   \kern-.3em \raise.4ex \hbox{$>$}}}}
\def\lsim{\mathrel{\hbox{\rlap{\lower.55ex \hbox {$\sim$}}
                   \kern-.3em \raise.4ex \hbox{$<$}}}}

\def\grb{GRB\thinspace{981226}}
\def\vla{VLA\thinspace{232937.2$-$235553}}
\def\ts{\thinspace}

\makeatletter
\renewenvironment{deluxetable}[1]{\def\pt@format{\string#1}%
\set@tblnotetext\global\pt@ncol=0\global\pt@column=0\global\pt@page=1%
\def\pt@addcol{\global\advance\pt@ncol by\@ne}}%
{\pt@width\wd\pt@box\box\pt@box\vskip-0.5cm\spew@ptblnotes%
\typeout{Page \the\pt@page\space of table \thetable\space has been set to
width \the\pt@width\space with \the\pt@nlines\space lines per page}%
\endcenter\vskip-0.5cm\end@float}
\def\startdata{\pt@line=0\pt@calcnlines%
\ifdim\pt@width>\z@\def\@halignto{to \pt@width}\else\def\@halignto{}\fi%
\let\fnum@table=\fnum@ptable\set@mkcaption%
\@float{table}[t]\center\caption{\pt@caption}\leavevmode%
\setbox\pt@box=\pt@tabular{\pt@format}\pt@head}
\def\thebibliography{\subsection*{REFERENCES}
\list{}{\labelwidth3em\leftmargin\labelwidth\labelsep\z@\parsep\z@
\itemsep\z@\itemindent-3em\usecounter{enumi}}
\def\refpar{\relax}
\def\newblock{\hskip .11em plus .33em minus .07em}
\sloppy\clubpenalty4000\widowpenalty4000
\sfcode`\.=1000\relax}
\makeatother

\lefthead{Frail et al. 1999}
\righthead{Radio afterglow of GRB 981226}
\begin{document}
 
 
\title{\large \bf The Radio Afterglow and the Host Galaxy of the X-Ray Rich
  GRB\thinspace{981226}}
 
\author
{D. A. Frail\altaffilmark{1},
 S. R. Kulkarni\altaffilmark{2},
 J.~S. Bloom\altaffilmark{2},
 S. G. Djorgovski\altaffilmark{2},
 V. Gorjian\altaffilmark{2},
 R. R. Gal\altaffilmark{2},
 J. Meltzer\altaffilmark{2}, 
 R. Sari\altaffilmark{3},
 F. H. Chaffee\altaffilmark{4},
 R. Goodrich\altaffilmark{4},
 F. Frontera\altaffilmark{5,6} \&\
 E. Costa\altaffilmark{7} 
}

\altaffiltext{1}
{National Radio Astronomy Observatory, P.~O.~Box O,
 Socorro, NM 87801.}

\altaffiltext{2}{Department of Astronomy, California Institute of
  Technology, MS 105-24, Pasadena, CA 91125.}

\altaffiltext{3}{Theoretical Astrophysics, California Institute of
  Technology, MS 103-33, Pasadena, CA 91125.}

\altaffiltext{4}
{W. M. Keck Observatory, 65-0120 Mamalahoa Highway, Kamuela, HI 96743.}

\altaffiltext{5}{Istituto de Technologie e Studio Radiozioni Extraterrestri
CNR, via Gobetti 101, I-40129 Bologna, Italy.}

\altaffiltext{6}{Dipartimento di Fisica, Universit\`a Ferrara, via
Paradiso 12, I-44100 Ferrara, Italy.} 

\altaffiltext{7}
{Istituto di Astrofisica Spaziale, CNR,
via Fosso del Cavaliere, Roma I-00133, Italy.}

\begin{abstract}

  We report the discovery of a radio transient \vla, coincident with
  the proposed X-ray afterglow for the gamma-ray burst \grb. This GRB
  has the highest ratio of X--ray to $\gamma$-ray fluence of all the
  GRBs detected by {\em BeppoSAX} so far and yet no corresponding
  optical transient was detected.  The radio light curve of \vla\ is
  qualitatively similar to that of several other radio afterglows.  At
  the sub-arcsecond position provided by the radio detection, optical
  imaging reveals an extended R=24.9 mag object, which we identify as
  the host galaxy of \grb.  Afterglow models which invoke a jet-like
  geometry for the outflow or require an ambient medium with a radial
  density dependence, such as that produced by a wind from a massive
  star, are both consistent with the radio data. Furthermore, we show
  that the observed properties of the radio afterglow can explain the
  absence of an optical transient without the need for large
  extinction local to the GRB.

\end{abstract}


\section{Introduction \label{sec:introduction}}

\grb\ was detected by the Gamma-ray burst monitor and the Wide Field
Camera (WFC) on the {\em BeppoSAX} satellite on 1998 December 26.41 UT
(Di Ciolo et al. 1998). The 6\arcmin\ error radius of the WFC was
further refined to 1\arcmin\ by the {\em BeppoSAX} Narrow Field
Instruments (NFI circle) which observed a fading X-ray transient 11 hours
after the burst (Frontera et al. 1998).  The primary interest in this
burst is its apparent X-ray richness: the burst has the highest X-ray
to $\gamma$-ray fluence ratio (Frontera et al. 1999) to date.

The 1-arcmin NFI error circle was intensively followed up by nearly
half a dozen telescopes around the world.  However, none of the
proposed candidates have been reliably established to be the optical
afterglow of the burst.  Here we report radio observations of the NFI
error circle.  We have identified a transient radio source with
properties similar to previously studied radio afterglows.  We propose
this to be the radio afterglow of \grb. With subsequent optical
observations we have identified a galaxy at the position of the radio
source, presumably the host galaxy of \grb.

\section{Observations \label{sec:observations}}

Radio observations began at the Very Large Array (VLA) on 1998
December 27.0 UT, 14 hr after the burst. At that time only a
preliminary WFC position was available (Piro 1998) and we imaged, in
the 4.86-GHz band, the entire 8-arcmin WFC radius error circle using a
four-pointing mosaic; see Table 1.  When the more accurate NFI
position was released (Frontera et al. 1998) we began to image in the
8.46 GHz band.  Synchrotron self-absorption is important at early
times and for this reason it is advantageous to conduct the initial
observations at higher frequencies (Shepherd et al. 1998).  However,
offsetting this advantage is the fact that the field-of-view is
inversely proportional to the frequency; for the VLA the
FWHM=45\arcmin/$\nu$(GHz) or 5.3\arcmin\ at 8.46 GHz.

All observations were performed using the VLA in its standard
continuum mode. At each frequency the full 100 MHz bandwidth was
obtained in two adjacent 50-MHz bands. The flux density scale was tied
to the extragalactic sources 3C\thinspace{48} (J0137+331) or
3C\thinspace{147} (J0542+498), while the array phase was monitored by
switching between the GRB and a VLA phase calibrator J2333$-$237. Data
calibration and imaging were carried out with the {\em AIPS} software
package.

In our image of 1999 January 3.95 UT we found an unresolved source at
high significance (6$\sigma$); see Figure 1. The source located at
(epoch J2000) $\alpha$\ =\ $23^h29^m37.21^s$ ($\pm{0.03^s}$) $\delta$\ 
=\ $-23^\circ55^\prime53.8^{\prime\prime}$ ($\pm{0.4}^{\prime\prime}$)
is well within the NFI error circle; we refer to this source as \vla.
In Figure 2 we present a light curve of the flux density measured at
this position.  For comparison, we present the light curve of another
source, VLA J232940.0$-$235254, located well outside the NFI circle.
From Figure 2 we see that \vla\ is clearly a variable source.
Quantitatively, this is revealed in the $\chi^2$ obtained under the
assumption of a constant flux density: the reduced $\chi^2$ of \vla\ 
is 5.7 whereas it is 0.7 for VLA J232940.0$-$235254.

\section{Discovery of the Radio Afterglow\label{sec:radio-afterglow}}

The radio light curve of \vla\ is qualitatively similar to previous
radio afterglows, most notably GRB 980703 (Frail et al.~1999).
Although GRB 980703 was an order of magnitude brighter, it too showed
a rise to maximum about 10 days after the burst, followed by a
power-law decay to a level where it was no longer visible. It is on
the basis of these similarities and the discovery of a underlying
galaxy (see \S{\ref{sec:host-galaxy}}) that we make the claim that
\vla\ is the radio afterglow of \grb. 

An unlikely (but not implausible) hypothesis is that \vla\ is an
active galactic nucleus (AGN). Indeed, if later observations show
detectable emission from \vla\ (and not attributed to emission from
the disk of the host galaxy) then this hypothesis would be favored.
In contrast, no late time re-brightening is expected in the radio
afterglow hypothesis. 

We now proceed with the hypothesis that \vla\ is the afterglow of
\grb. The qualitative behavior of other radio afterglows can be
summarized as follows: the flux at a given radio frequency ($\nu_R$)
rises as a power law $F_R(t) \propto t^{\alpha_r}$, reaches a broad
maximum at $F_m$ and then decays as $F_R(t)\propto t^{\alpha_d}$.  The
epoch of peak flux is frequency dependent, $t_m(\nu_R)$. In
anticipation of our discussion in \S{\ref{sec:discussion}} we have
carried out a fit to this model. Given the paucity of data we fixed
$t_m({\rm 8.46~GHz})=8.5$ d, a value that is reasonable (Figure 2),
although somewhat smaller than those of previous radio afterglows
(e.g. Taylor et al. 1998, Frail et al. 1999).  We obtain the following
model $\alpha_r=+0.8\pm{0.4}$, F$_m(\rm 8.46 GHz)=173\pm{27}$ $\mu$Jy and
$\alpha_d=-2.0\pm{0.4}$ (overall $\chi^2=9$ with 8 degrees of
freedom). The errors quoted on each parameter are 1$\sigma$ assuming
all other parameters are fixed; the covariance between parameters is
not reflected in these errors. Choosing $t_m$ a few days on either
side of our value does not substantially change the fitted parameters.
Parenthetically, we note that at $b=-71^\circ$ the path length \grb\ 
takes through the turbulent ionized gas of our Galaxy is small and
therefore strong modulation of the afterglow flux by interstellar
scattering is not expected.


\section{Host Galaxy of GRB 981226 \label{sec:host-galaxy}}

We observed the radio transient position with the Low Resolution
Imaging Spectrometer (Oke et al. 1995) at the Keck II telescope on
Mauna Kea on three nights in June 1999: June 11 UT (SGD, B. A. Jacoby),
June 19 UT (B. Schaeffer, FC) and June 21 UT (L. Hillenbrand). In each
case we obtained images in the R band with total integration time of
1400 s, 720 s and 3360 s respectively.  After correcting for the bias
and pixel variation (flat fielding) we registered the images and
combined  them to form the final image.

Using the USNO A2.0 catalog (Monet et al.  1998) we computed an
astrometric plate solution to one of the images of June 11.  The
r.m.s.~statistical uncertainty of the solution, computed from the 12
tie stars, is $\sigma_{RA}$=0.32 and $\sigma_\delta$=0.17 arcsec.  To
obtain the total uncertainty relative to the radio transient position,
we add these uncertainties in quadrature with the empirical
uncertainty of $\sigma_{RA}$=0.17 and $\sigma_\delta$=0.18 of the
USNO-A2.0 astrometric tie to the International Celestial Reference
Frame (Deutsch 1999). The position of the host galaxy (discussed
below) is quoted with respect to the astrometry of this image.
The photometric zero point was determined by observations of
the standard star field Mark A (Landolt 1992) on 21 June 1999 which
was a photometric night.  In order to facilitate comparison of our
photometry we note that star ``A'' at R.A.=
23:29:35.6 and $\delta=-$23:55:38.0 (J2000) has R = 20.81 $\pm 0.02$.

In Figure 3 we show a $50 \times 50$ arcsec optical region centered on
the position of \vla. We note an extended object, presumably a galaxy,
0.55 arcsec West and 0.41 arcsec South of \vla. We propose this object
to be the host galaxy of \grb.  Despite the offset, the position of
the radio transient and optical galaxy are consistent within the
astrometric uncertainties.  While there is visually an indication of
extension beyond the seeing FWHM (=0.85 arcsec), the signal-to-noise
in our Keck image is not sufficient to reliably produce a measure of
the half-light radius. We find the magnitude of the putative host to
be R = 24.85 $\pm 0.06$ mag in an 1.1 arcsec aperture radius about the
centroid.  Uncertainties arising from the unknown color term (images
were obtained in R-band only) have not been included. Both the
apparent magnitude and the small extension for the proposed host are
completely consistent with normal galaxies at $z\sim$1 and beyond (see for
e.g. Hogg \&\ Fruchter 1999, Mao \&\ Mo 1999).

\section{No Detectable Optical Afterglow\label{sec:no-optical}}

Searches for the optical afterglow emission were carried out by a
number of groups independent of our radio effort and results reported
in the GCN\footnotemark\footnotetext{The GRB Coordinates Network (GCN)
  is a service to the GRB community run by Scott Barthelmy \hfil\break
  and Paul Butterworth and can be accessed at \hfil\break
  $\rm{http://lheawww.gsfc.nasa.gov/docs/gamcosray/legr/bacodine/gcn\_main.html.}$}.
Given the important diagnostic value of multi-wavelength observations
we now summarize the results of the optical efforts and then proceed
in the next section to use afterglow theory to see if the existing
optical upper limits and radio light curve provide any significant
constraints.

No fewer than three candidates were proposed as the optical afterglow
from \grb\ and their sky location can be found in Figure 1. Galama et
al.~(1998) reported a source within the NFI error circle that was seen
in their R-band images but was not visible in the Digitized Sky Survey
(DSS) taken in the UK Schmidt red filter.  Follow-up R-band
observations (e.g.  Rhoads et al. 1998, Bloom et al.  1998, Schaefer
et al. 1998) found no evidence for variability. The absence of the
source in the DSS is likely due to the different filters employed for
the two comparison images.

Castro-Tirado et al.~(1998) proposed a second candidate in the NFI error
circle. From preliminary photometry of J-band images taken at the Calar
Alto 3.5-m they found evidence for variability between 1998 December
26.76 UT (J=19.4) and 1998 December 27.76 UT (J=20.5).  Wozniak (1998)
and Lindgren et al.~(1998) did detect the object in the I and R bands,
respectively, but they find no evidence for variability.

A third candidate identified by Wozniak (1998) appears to be a genuine
variable (Bloom et al. 1998) but it lies well outside the NFI error
circle (but inside the WFC) and thus its relationship to \grb\ seems
remote.

The best upper limits to a possible optical transient were obtained by
Lindgren et al.~(1999) with data obtained at the 2.5-m Nordic Optical
Telescope and the 1.5-m Danish Telescope.  They report no evidence for
significant variability in the NFI error circle down to a limiting
magnitude of R$\sim$23 mag for images taken at 0.4, 1.4 days and 2.4
days after the burst.  The correction for Galactic extinction in this
direction $(l=38^\circ, b=-71^\circ)$ is small (A$_R$=0.06).

\section {Discussion \label{sec:discussion}}

The primary interest in \grb\ is its X-ray richness (Frontera et al.
1999).  This burst of 20-s duration was otherwise unremarkable
(DiCiolo et al. 1998). In the currently accepted picture for GRBs, the
initial burst is due to internal shocks of relativistically moving
material ejected by a central object.  The afterglow emission arises
as the relativistic ejecta is slowed down by the surrounding gas.
From an afterglow perspective, the most interesting aspect of this GRB
is the short-lived radio afterglow.  The time to rise to the peak flux
of about 8 d is typical of almost all previously studied radio
afterglows. The short duration of the afterglow is therefore mainly
due to the fast decline.  Below we discuss a number of afterglow
models which can account for this rapid decline in the radio flux

We begin with the simplest model: a spherical burst expanding into a
constant density medium (Sari, Piran \& Narayan 1998). In this case
the fits in \S{\ref{sec:radio-afterglow}} predict that the optical
flux would have reached the maximum value of $F_m$ (corresponding to
roughly 18 mag) at epoch $t_O=t_m({\rm 8.46~GHz})
(\nu_R/\nu_O)^{-2/3}\simeq{500}$ s; here $\nu_R=8.46$ GHz, and
$\nu_O=5\times 10^{14}$ Hz. The expected flux at later times would
then be $F_m(t/t_O)^{\alpha_d}$.  In this framework we would expect no
detectable optical flux at the epoch of the first deep R-band
observations, $t=0.4$ hr (Lindgren et al. 1999) unless
$\alpha_d>-1.1$.  Our fits to the radio light curve rule out such a
shallow decay at the 2.3$\sigma$ level, and thus the observations are
consistent with this model without invoking extinction. However, a
logical conclusion of this model is that the underlying power law
index of the shocked electrons (in the forward shock) is then
$p=1-4/3\alpha_d\sim 3.7\pm 0.5$ much higher than the 2.2--2.5 value
inferred in most bursts (Sari, Piran \& Halpern 1999). For this reason
we are not in favor of this model.

Two models have been proposed to explain rapid fading at X-ray and 
optical wavelengths: jets (Rhoads 1999, Sari et al. 1999) and
expansion into an ambient medium with a radial density dependence
($\rho\propto{r}^{-2}$), such as that produced by a wind from a
massive star (Chevalier \&\ Li 1999; see also Vietri 1997).
Both these models are attractive because they can also account for
the fast decline of the radio emission.

First we interpret the data presented here in the framework of the jet
model as presented for the optical-radio data of GRB 990510 by
Harrison et al. (1999). 

In the jet model radio emission above the synchrotron self-absorption
frequency is supposed to rise as $t^{1/2}$, reach a maximum and then,
for epoch $t>t_{\rm jet}$, start a gentle decline as $t^{-1/3}$; here
$t_{jet}$ is the epoch when the bulk Lorentz factor of the ejecta
becomes comparable to the inverse of the opening angle of the jet.  At
a later epoch when the maximum frequency $\nu_m$ falls below the
observing frequency, the radio flux is expected to plummet as
$t^{-p}$; see Harrison et al. (1999) for a brief review and
application to GRB 990510.  The radio light curve is consistent with
this picture if one uses parameters similar to that of GRB\ts{990510},
$t_{jet}\lsim{t_m}({\rm 8.46~GHz}) \sim 8.5\ {\rm d}$ and $p\sim 2.5$.
The value of $t_{jet}$ is not well constrained but the rise in the
flux at early times as $\alpha_r=+0.8\pm{0.4}$ (consistent with
spherical expansion) suggests that $t_{jet}>5$ d.

The evolution of the afterglow emission at early times when the
optical data were taken ($t<t_{\rm jet}$) is indistinguishable from
the spherical case detailed above, except that in the jet model the
observed maximum in the radio flux of $173\pm 27\,\mu$Jy is about a
factor of two below the optical maximum at $t_O$.  Here again, if we
adopt a simple decay with $\alpha_d\simeq-1.2$ we find the early decay
of the light curve is sufficient to render the optical afterglow
undetectable at the epoch of the first measurement without requiring
extinction within the host galaxy.

In the case of the wind-shaped circumstellar medium model, the
afterglow emission is weakened as the relativistic blast wave ploughs
into ambient material with decreasing density.  Chevalier \&\ Li
(1999) have applied this model to the afterglow of GRB 980519. We find
that for reasonable stellar wind parameters this model can fit the
light curve in Figure 2. Extrapolating this model into the optical
band at t=0.4 day (the epoch of the first observations by Lindgren et
al. (1999)) we predict F$_m$=870 $\mu$Jy at $\nu_m=8\times{10}^{11}$
Hz, $p=3$ and the cooling frequency $nu_c=8\times{10}^{16}$ Hz.  The
expected R-band flux density in this model, given by
F$_R$=F$_m(\nu_R\nu_m)^{-(p-1)/2}$, lies below the observed magnitude
limit.

In summary, regardless of which model we use (spherical, jet,
circumstellar), we find that we can explain the radio afterglow
emission without invoking extinction to explain the absence of the
optical afterglow. While we cannot rule out large extinction local to
the GRB, it is not required. Unfortunately, the paucity of the data
does not allow us to chose uniquely the two competing models.  If this
GRB was the end product of a massive star then the short duration of
the radio afterglow could well be due to the blast wave running out of
the dense circumstellar medium. Indeed, as noted from radio studies of
type II SNe (surely the end products of massive stars), there is
considerable structure in the distribution of circumstellar matter.
Radio emission from a number of SNe shows undulations which can be
ascribed to successive rings of material (Weiler et al. 1992) and in
other cases the sudden cessation of radio emission (including 1987A)
can be due to an edge in the circumstellar matter (Staveley-Smith et
al. 1992, Montes et al. 1998).  The above discussion highlights the
growing and unique contributions that radio afterglow can make to the
study of GRBs. In the future, high signal-to-noise radio observations
of GRBs should be capable of showing whether the emission simply
peters out or is reduced abruptly.  If the latter then radio afterglow
will allow us to probe the structure of the circum-burst medium.

\acknowledgements

SRK's research is supported by grants from NSF and NASA. We thank B.
Schaefer, B. Jacoby, L. Hillenbrand and Chelminiak for making some of
the optical observations. The Very Large Array (VLA) is operated by
the National Radio Astronomy Observatory which is a facility of the
National Science Foundation operated under a cooperative agreement by
Associated Universities, Inc.


\def\pni{\par\noindent}
\bigskip
\centerline{\bf REFERENCES}

\pni Bloom, J. S., Gal, R. R., \& Meltzer, J. 1998, GCN 182

\pni    Castro-Tirado et al. 1998,
        GCN 172        
        
\pni    Chevalier, R. A., \& Li, Z. -Y. 1999,
        ApJL submitted, astro-ph/9904417
        
\pni    Deutsch, E. W. 1999,
        astro-ph/9906177

\pni    Di Ciolo, L., Celidonio, G., Gandolfi, G., in't Zand, J. J., Heise, J.,
        Costa, E., \& Amati, L. 1998, IAUC 7074
        
\pni    Frail, D. A. et al. 1999,
        in preparation    
        
\pni    Frontera, F. et al. 1998,
        GCN 184

\pni    Frontera, F. et al. 1999,
        in preparation    
        
\pni    Galama, T. et al. 1998,
        GCN 172

\pni    Harrison, F. A. et al. 1999, ApJ, submitted, astro-ph/9905306

\pni    Hogg, D. W., \& Fruchter, A. S. 1999,
        ApJ in press, astro-ph/9807262

\pni    Landolt, A. U. 1992,
        AJ, 104, 340

\pni    Lindgren, B. et al. 1998,
        GCN 190

\pni    Mao, S. \&\ Mo, H. J. 1999,
        A\&A, 339, L1

\pni    Monet, D. et al. 1998,
        USNO-A2.0: A Catalog of Astrometric Standards, U.S.
        Naval Observatory.

\pni    Montes, M. J., Van Dyk, S. D., Weiler, K. W., Sramek, R. A., Panagia,
        N. 1998 ApJ, 506, 874

\pni    Oke, J. B. et al. 1995, PASP, 107, 375.

\pni    Piro, L. 1998,
        GCN 174
        
\pni    Rhoads, J. E., Orosz, J. A., Lee, J., \& Stassun, K. 1998
        GCN 181
        
\pni    Rhoads, J. E. 1999,
        ApJ submitted, astro-ph/9903399
        
\pni    Sari, R., Piran, T., \& Narayan, R.~1998, ApJ, 497, L17

\pni    Sari, R., Piran T. \& Halpern, J. P. 1999, 
        ApJ, 519, L17

\pni    Schaefer, B. E., Kemp, J., Feygina, I., \& Halpern, J. 1998, 
        GCN 185

\pni    Shepherd, D. S., Frail, D. A., Kulkarni, S. R. \&\
        Metzger, M. R. 1998,
        ApJ, 497, 859

\pni    Staveley-Smith, L. et al. 1992,
        Nature, 366, 166

\pni    Taylor, G. B., Frail, D. A., Kulkarni, S. R.,
        Shepherd, D. S., Feroci, M. \&\ Frontera, F. 1998,
        ApJ, 502, L11

\pni    Vietri, M. 1997, ApJ, 488, L105

\pni    Weiler, K. W., Van Dyk, S. D., Pringle, J. E., \&  Panagia,
        N. 1992, ApJ, 399, 672

\pni    Wozniak, P. R. 1998,
        GCN 177
        

\newpage

\begin{deluxetable}{lrcrr}
\tabcolsep0in\footnotesize
\tablewidth{\hsize}
\tablecaption{VLA Observations of \grb}
\tablehead {
\colhead {Epoch }      &
\colhead {$t-t_0$ } &
\colhead {$\nu_{obs}$} &
\colhead {$S_{RT}\pm\sigma$} & 
\colhead {$S_{field}\pm\sigma$} \\
\colhead {(UT)}      &
\colhead {(days)} &
\colhead {(GHz)} &
\colhead {$(\mu{\rm Jy})$} & 
\colhead {$(\mu{\rm Jy})$} 
}
\startdata
1998 Dec. 27.00 & 0.59  & 4.86 & 105$\pm$48      & \omit         \nl
1998 Dec. 29.92 & 3.51  & 8.46 & 73$\pm$27       & 136$\pm$27    \nl
1998 Dec. 30.95 & 4.54  & 8.46 & 143$\pm$45      & \omit         \nl
1999 Jan. 03.95 & 8.54  & 8.46 & 169$\pm$28      & 114$\pm$28    \nl
1999 Jan. 07.98 & 12.57 & 4.86 & $-$13$\pm$43    & \omit         \nl
1999 Jan. 07.98 & 12.57 & 8.46 & 67$\pm$29       & 178$\pm$29    \nl
1999 Jan. 11.06 & 15.65 & 8.46 & 80$\pm$30       & 140$\pm$30    \nl
1999 Jan. 16.09 & 20.68 & 8.46 & $-$28$\pm$30    & 153$\pm$30    \nl
1999 Jan. 19.89 & 24.48 & 8.46 & 27$\pm$14       & 145$\pm$14    \nl
1999 Jan. 21.85 & 26.44 & 8.46 & 37$\pm$20       & 103$\pm$20    \nl
1999 Mar. 04.72 & 68.31 & 8.46 & $-$14$\pm$19    & 132$\pm$19    \nl
1999 Mar. 05.93 & 69.52 & 8.46 & $-$14$\pm$14    & 142$\pm$14    \nl
1999 Mar. 28.74 & 92.33 & 8.46 & 9$\pm$25        & 111$\pm$25    \nl
1999 May  27.44 & 152.03 & 8.46 & 15$\pm$20        & 133$\pm$20    \nl
\enddata
\tablecomments{The columns are (left to right), (1) UT date of the
  start of each observation, (2) Time elapsed in days since the GRB
  981226 event, (3) The observing frequency, (4) The peak flux density
  at the position of the radio transient with the error, given as the 
  root mean square flux density, and (5) The peak flux density of a comparison
  source in the field. The synthesized beam at 8.46 GHz is of order 4.5\arcsec\
  between 1998 December and 1999 January, 
  while for observations in 1999 March the beamsize is 12\arcsec. The beamsize
  at 4.86 GHz was approximately 6.7\arcsec.}
\end{deluxetable}

\clearpage
\begin{figure*}
\vskip-1truein
\centerline{\hbox{\psfig{figure=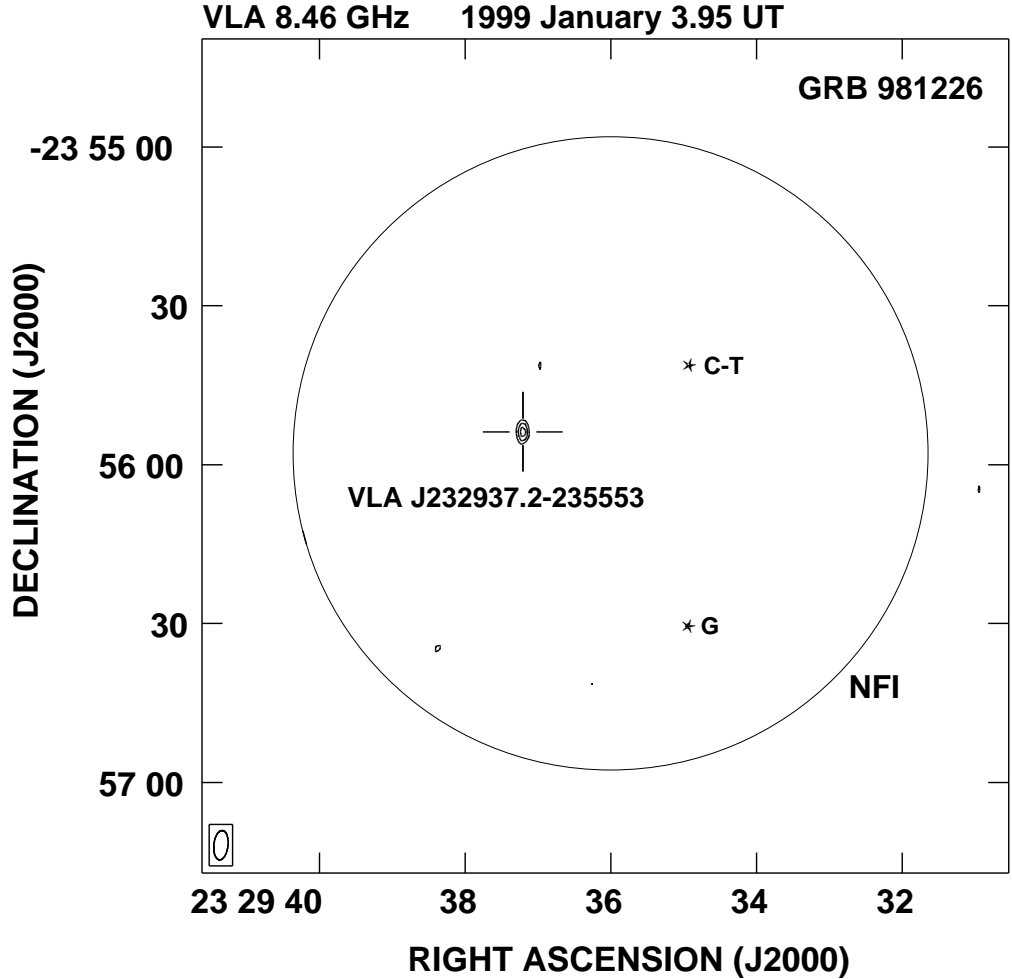,width=6.0in,angle=0}}}
\caption[]{An image of the field of the radio transient of
  GRB 981226, taken with the VLA at 8.46 GHz on 1999 January 3.95 UT.
  The arcminute error radius of the X-ray afterglow as detected by the
  NFI of {\em BeppoSAX} is indicated by the large circle. The radio
  transient (VLA J232937.2$-$235553) lies between at the center of the
  cross.  Contours are plotted in steps of 3.5, 4.5 and 5.5 times the
  rms noise of 28 $\mu$Jy/beam. The shape of the
  5.6\arcsec$\times$2.7\arcsec\ beam in shown is the lower left
  corner.  The two asterix symbols indicate the position of the
  optical afterglow candidates proposed by Galama et al. (G) and
  Castro-Tirado et al. (C-T). See text for more details.
\label{fig:jan03-radio}}
\end{figure*}

\clearpage
\begin{figure*}
  \centerline{\hbox{\psfig{figure=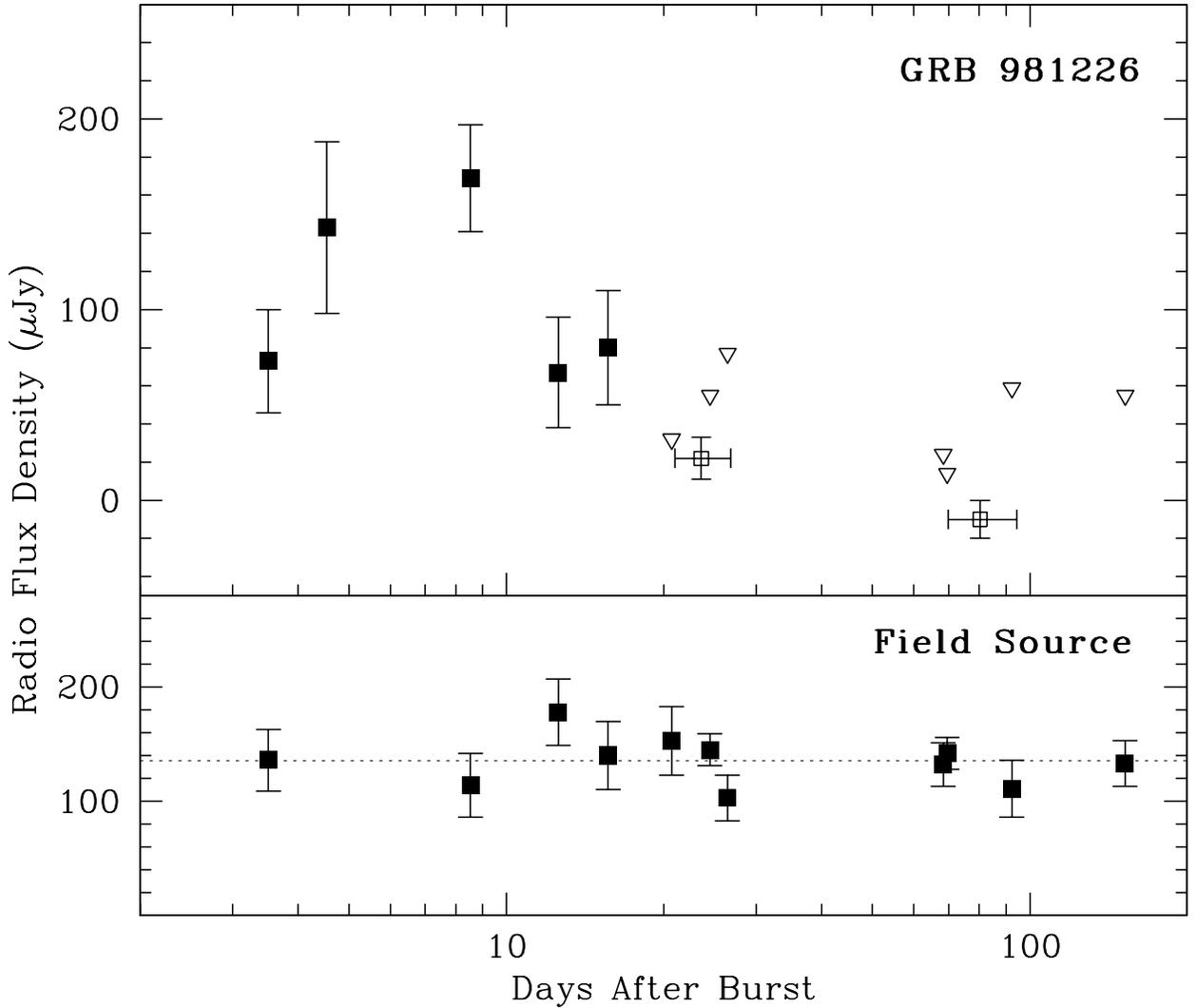,width=7.0in}}}
\caption[]{A 8.46 GHz light curve of the radio transient 
  of \grb\ (top panel), and a background source in the field (bottom
  panel).  Detections at each epoch are indicated by the filled
  squares.  Non-detections are given the open triangles, defined as
  the peak brightness at the location of VLA J232937.2$-$235553 plus
  two times the rms noise in the image.  The peak brightness for
  weighted averages of three adjacent epochs are shown by open
  squares.
\label{fig:rlightc}}
\end{figure*}

\clearpage
\begin{figure*}
\centerline{\hbox{\psfig{figure=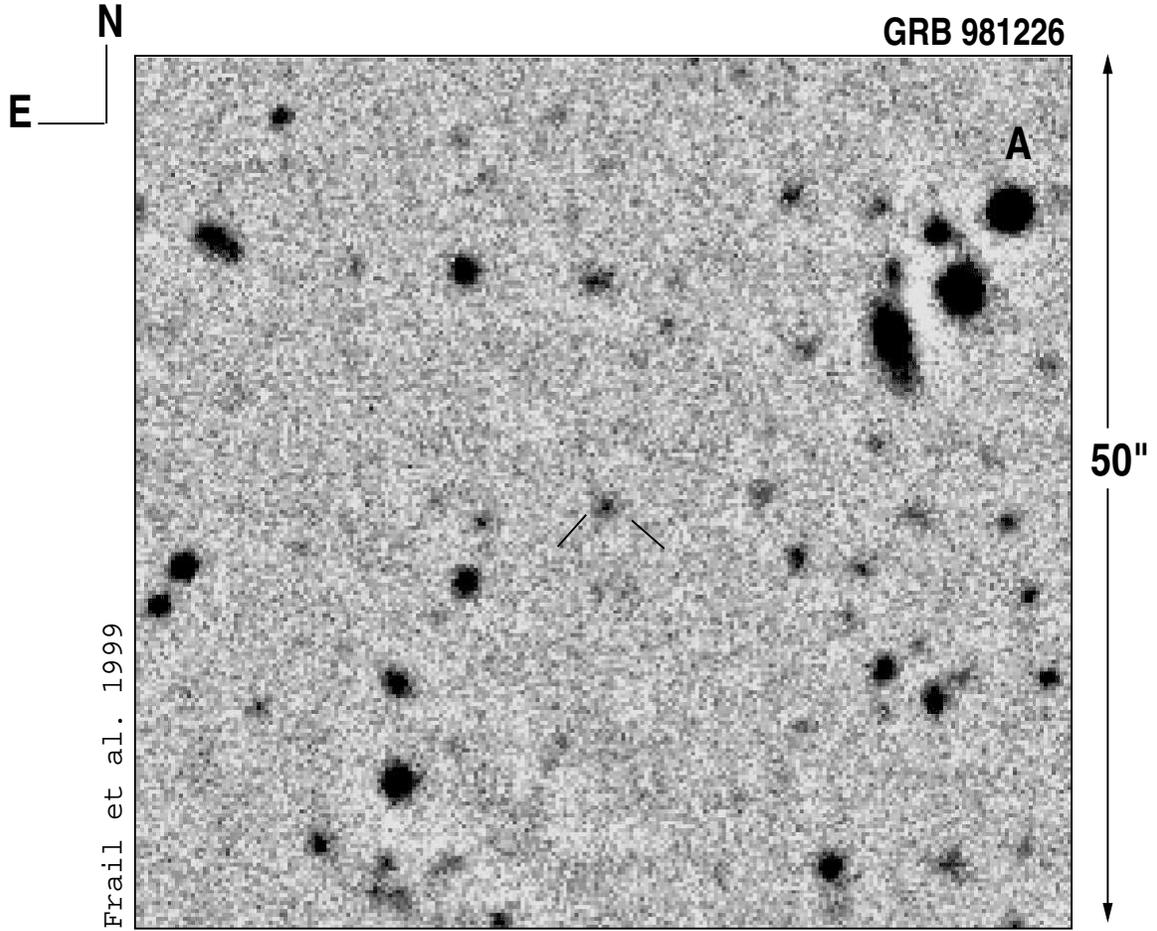,width=6.0in,angle=270}}}
\caption[]{Deep optical image of the field centered on the radio afterglow of
  GRB 981226.  This 3360-s R-band exposure on the Keck II 10-m LRIS
  (Oke et al. 1995) instrument reveals a faint galaxy coincident
  (within the astrometric errors) with R = 24.85 $\pm 0.06$.  Star "A"
  is labeled as a photometric reference (see text).
\label{fig:optical-fig}}
\end{figure*}

\end{document}